# High-Quality Planar High-$T_c$ Josephson Junctions

N. Bergeal[a)], X. Grison, J. Lesueur

*Laboratoire de Physique Quantique, ESPCI – UPR5 CNRS, 10 rue Vauquelin, 75231 Paris (France)*

G. Faini

*Laboratoire de Photonique et Nanostructures du CNRS, Route de Nozay, 91460 Marcoussis (France)*

M. Aprili[b)]

*CSNSM, CNRS-IN2P3, Bât 108, 91405 Orsay (France)*

J. P. Contour

*Unité Mixte de Physique CNRS/THALES, Domaine de Corbeville, 91404 Orsay (France)*

**Abstract**

Reproducible high-$T_c$ Josephson junctions have been made in a rather simple two-step process using ion irradiation. A microbridge (1 to 5 µm wide) is firstly designed by ion irradiating a c-axis-oriented $YBa_2Cu_3O_{7-\delta}$ film through a gold mask such as the non-protected part becomes insulating. A lower $T_c$ part is then defined within the bridge by irradiating with a much lower fluence through a narrow slit (20 nm) opened in a standard electronic photoresist. These planar junctions, whose settings can be finely tuned, exhibit reproducible and nearly ideal Josephson characteristics. This process can be used to produce complex Josephson circuits.

a) Electronic mail : nicolas.bergeal@espci.fr

b) Also at : *Laboratoire de Physique Quantique, ESPCI – UPR5 CNRS, 10 rue Vauquelin, 75231 Paris (France)*

For the last fifteen years or so, a great deal of research have been devoted to High-$T_c$ Superconducting (HTSc) Josephson Junctions (JJ) for electronic applications [1]. Two main techniques have been used : the so called Grain Boundary junctions (GBj) and the edge junctions. Both of them require special substrates, which strongly limit the design of complex circuits, and delicate processing, which impairs their reproducibility, and therefore their applications. An alternative process has been proposed to create weak links on purpose. The atomic disorder induced by ion irradiation drives HTSc cuprates towards an insulating state ; for intermediate disorder, the $T_c$ is lowered and the resistivity is increased [2,3]. Authors already succeeded in making Superconductor-Normal-Superconductor (SNS) junctions by lowering locally the $T_c$ of a HTSc film[4]. They obtained rather good Josephson characteristics, but the overall process was rather complex still.

Here, we present an original method to make HTSc JJ. A t=150 nm thick c-axis oriented $YBa_2Cu_3O_{7-\delta}$ (YBCO) film is pulsed laser deposited and in-situ covered by a 40 nm thick gold layer to insure both low contact resistances to the junction and reproducible characteristics. PMMA photoresist is then deposited and patterned to design microbridges, (1 to 5 µm wide, 8 to 40 µm long). A 250 nm thick gold layer is deposited and lifted-off, such as the microbridges and the contacts remains covered. The in-situ gold layer is removed by Ar Ion Beam Etching. The sample is then ion-irradiated with 110 keV Oxygen ; the $5\times10^{15}$ at/cm$^2$ fluence makes the unprotected parts insulating, therefore designing a current path underneath the gold layer including the bridges and the contact pads. No HTSc material is removed during this process. In a second step, gold above the microbridge is removed by Ar IBE through a suitable polymethyl methacrylate (PMMA) mask. Finally, photoresist is deposited all over the sample, and a narrow slit (20nm width) is opened across the microbridge which will defined the junction area. 100 keV Oxygen ions are used to lower the $T_c$ in this region, with typical fluences of a few $10^{13}$ at/cm$^2$. We therefore end with a junction completely embedded in a cuprate layer, contacted with in-situ low resistance gold pads. No further annealing or heat treatment is needed to obtain the characteristics described below, as opposed to previous studies[5].

In the inset (a) Figure 3 is displayed the resistance as a function of temperature for two 1 µm width bridges irradiated with 3 and $6\times10^{13}$ at/cm$^2$ , and measured with a tiny current : the highest transition refers to that of the electrodes (the same as the unprocessed film) , and the lowest, which can be tuned by the fluence, to the junctions themselves. The resistances of the bridges are the same within 10% , which gives an order of magnitude of their geometrical dispersion. The transitions being sharp (ΔT = 1

– 2 K), one can adjust precisely the operation temperature of the device. Let us emphasize that the observed transition corresponds to the Josephson coupling $T_J$ of the two electrodes and *not* the transition of the irradiated part itself. As a matter of fact, the inset Figure 1 shows the I-V characteristics of a junction at different temperatures. At high temperature, just below the observed zero resistance state in the inset (a) Figure 3, a typical RSJ like behavior is observed [6] with an upward curvature of the dissipative branch, indicating a Josephson coupling of the two electrodes through a normal layer. When lowering the temperature, the I-V characteristics displays a downward curvature as expected for a flux flow regime. The temperature which separates the two regimes is $T_c$', the critical temperature of the irradiated part (32 (±1) K in this case).

The critical current of the junctions $I_c(T)$ shows a quadratic dependence as a function of temperature (solid lines in Figure 1). This is consistent with previous studies[5,7], and follows the deGennes-Wertammer model of proximity effect[8]. The coupling of the two pristine superconducting regions occurs through the damaged one above its own $T_c$'; Cooper pairs diffuse within a so called « normal coherence length » $\xi_N(T)$, corresponding to the lost of phase coherence due to thermal excitations :

$$\xi_N(T) = \left(\frac{\hbar D}{2\pi k_B T}\right)^{\frac{1}{2}} \left(1 + \frac{2}{\ln(T/T_c')}\right)^{\frac{1}{2}} \qquad (1)$$

where D is the diffusion constant. The critical current is therefore given by :

$$I_c(T) = I_0 \left(1 - \frac{T}{T_J}\right)^2 \frac{l/\xi_N}{\sinh(l/\xi_N)} \qquad (2)$$

where $I_0$ is a typical critical current, and *l* the length of the normal part assuming a rectangular shape in first approximation. This behavior is valid (and observed) for temperatures above $T_c$' of course. Since the temperature dependence of $\xi_N(T)$ itself is weak (sqrt(T) and log(T)), the main contribution to $I_c(T)$ comes from the divergence at $T_J$, and therefore follows a $T^2$ law. From the fitting procedure, one gets precisely $T_J$ whose dispersion is low (less than 0.5K for the $3\times10^{13}$ at/cm$^2$, and 3K for the $6\times10^{13}$ at/cm$^2$ junctions). The typical parameters just below $T_J$, where the device will be operated ($I_c$, $J_c$ (the critical current density) and the $I_cR_n$ product) are given in table I. These numbers compare nicely to the actual performances of low-$T_c$ junctions used for the RSFQ logic for example[9]. The high $J_c$ values combined with a low $R_n$ (insuring non-hysteretic behavior) look promising. Lastly, one can get from the fit the ratio $l/\xi_N(T_J)$, but the accuracy is poor since it enters in a slowly variating function (x/sh(x) at small x). These two lengths are important : $\xi_N(T_J)$ is governed by the physics of the junction, and *l* defines its geometrical properties. We will return to their determination later on.

The most stringent quality test for a JJ is the Fraunhofer pattern. In Figure 2 is shown the modulation of the critical current of a junction as a function of the applied magnetic field for different temperatures. The curve at T=73.4K is very close to |sinc(B)| as expected for a rectangular junction when its width W is smaller than the Josephson penetration length $\lambda_J = \left(\frac{\hbar}{2\mu_0 e}\right)^{1/2} \left(\frac{tW}{I_c(2\lambda_L + l)}\right)^{1/2}$ , where $\lambda_L$ is the London penetration depth. Lowering the temperature makes the critical current increase and $\lambda_J$ becoming smaller than W : the Owen and Scalapino model[10] taking into account self-field effects in the junction gives rise to typical patterns as observed here (no cancellation of $I_c$ and a less rounded pattern). Numerical comparisons of $\lambda_J$ and W in these experiments confirm nicely this cross-over (see legend Figure 2). Such an ideal behavior has not been reported in previous studies [4,5]. The period of the modulation corresponds to one flux quantum in the junction provided $\lambda_L$ is taken to be typical of thin films [11] (400 nm at 72K).

Besides the fact that we are able to make nice HTSc Josephson junctions in a rather simple process starting from blank c-axis oriented YBCO films, let us emphasize that we can tailor their characteristics rather precisely. Choosing the operating temperature is quite easy by choosing the fluence, but what about the exact geometry of the device ? It will control its electrical parameters as the normal state resistance and the critical current density. The geometry will be controlled firstly by the actual ion random walk within the sample, secondly by the modification of $T_c$ and the resistivity ρ of the material by defects, and thirdly by the proximity effect between the pristine region and the irradiated one. Using TRIM code[12] and suitable assumptions, we have computed a gaussian-like distribution of defects centered on the 20 nm wide slit whose standard deviation is roughly 80 nm [13], much smaller than a crude estimate of 150 nm or so[4], and in good agreement with experimental observations : the normal state resistance measured corresponds to the expected width of the channel within a few percents, even for the 1 μm wide ones, confirming that the straggling is much smaller than this value ; based on the actual R(T) curve of a pristine channel, we have been able to compute the one of irradiated samples with a very good accuracy (Figure 3). The increase in resistivity and the decrease in $T_c$ as a function of the created defects have been taken from the literature[3,14-16]. The resistance of the whole irradiated channel has been then calculated by integrating the resistivity over the defects distribution. As one can see in Figure 3, the overall shape is good but the low temperature part deviates from the experimental data : the estimated resistance is higher, and $T_c$' is lower. This is due to the proximity effect and the Josephson coupling. A better result for the resistance is obtained if one uses a 40 nm standard deviation defects distribution, i.e. a FWHM of 80 nm : *this is the actual length of the junction l*. The superconducting state from the reservoirs tends to propagate beyond the actual defects profile, and makes *l* shorter than crudely estimated. $T_J$ is much higher than $T_c$' due to Josephson coupling. The difference between computed and experimental data is controlled by the proximity effect, and the temperature dependence of the coherence length $\xi_N(T)$ within the irradiated region. A self consistent calculation is being made[13] to get a more physical insight in the proximity effect in such devices. As far as the applications are concerned, the precision of the above calculation shows that the overall process is under control, and the spread in characteristics low.

Reproducible HTSC junctions have been produced by ion-beam irradiation, whose characteristics can be adjusted precisely on a wide range of temperatures, and are suitable for many applications like SQUIDs and RSFQ logic, with a high degree of on-chip integration and complexity. A computation has been developed to model the junction geometry : the proximity effect makes the actual size of the device much smaller than a naive estimate from the defects distribution.

The authors gratefully acknowledge O. Kaitasov and S. Gautrot for the ion irradiation made at IRMA-CSNSM (Orsay-France), E. Jacquet, F. Lalu and L. Leroy for technical support.

# Captions

Figure 1 : Critical current as a function of temperature for irradiation doses 3 and $6\times10^{13}$ ions/cm$^2$ : two 5 µm wide samples are displayed for each, and the solid lines are quadratic fits (see text). Inset : Current vs Bias curves of a 5 µm wide bridge irradiated with $6\times10^{13}$ ions/cm$^2$ at different temperatures (T=29, 31, 32, 33, 36,40K from top to bottom). At high temperature, the upward curvature refers to RSJ-like behavior of a Josephson junction. At low temperature, the downward curvature is characteristic of a flux-flow regime. The arrow indicates $T_c$' estimated from this diagram.

Figure 2 : Critical current as a function of the applied magnetic field on a 5 µm wide junction irradiated at $3\times10^{13}$ ions/cm$^2$, for different temperatures. Taking the zero-temperature London penetration depth as 140 nm, $\lambda_J$ is calculated to be 1.9 µm, 2.9 µm and 4.2 µm for T=71.5 K, 72.5 K and 73.5 K respectively. Self-fields effects show-up when $\lambda_J$ becomes of the order of the junction width. The solid line is a fit to a perfect Fraunhofer pattern for a rectangular junction in the small junction limit. The insert is a schematic side view of the irradiation process.

Figure 3 : Resistance as a function of temperature for a 1 µm wide micro-bridge irradiated at $3\times10^{13}$ ions/cm$^2$, computed with a 80 nm long junction (dotted line) or a 40 nm long one (solid line). The absolute value of R(100K) has been adjusted within 5%. Inset (b) : blow-up of the main figure. Inset (a) : R(T) for two 1 µm wide bridges irradiated with 3 and $6\times10^{13}$ ions/cm$^2$ respectively. The Josephson behavior is observed between $T_c$' and $T_J$, shown here for the lower fluence.

**Table I**

Table I : Typical parameters for 5 μm wide junctions.

| Sample | Fluence | $T_j$ | $T_c'$ | $R_n$ (0.9$T_j$) | $I_c$(0.9$T_j$) | $I_cR_n$ (0.9Tj) | $J_c$ (0.9$T_j$) |
|---|---|---|---|---|---|---|---|
| M11 | 6×10$^{13}$ at/cm$^2$ | 49 K | 32 K | 1.2 Ω | 72 μA | 90 μV | 10 k A/cm$^2$ |
| M13 | 6×10$^{13}$ at/cm$^2$ | 48 K | 31 K | 1.2 Ω | 90 μA | 72 μV | 12 kA/ cm$^2$ |
| M21 | 3×10$^{13}$ at/cm$^2$ | 75 K | 57 K | 0.35 Ω | 772 μA | 270 μV | 100 kA/cm$^2$ |
| M25 | 3×10$^{13}$ at/cm$^2$ | 75 K | 61 K | 0.25 Ω | 1100 μA | 272 μV | 140 kA/cm$^2$ |

**Table I : N. Bergeal et al**

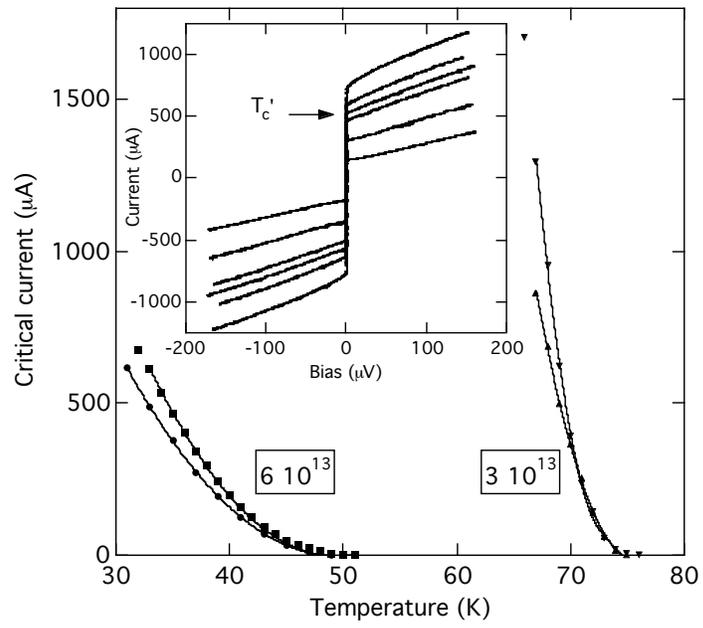

**Figure 1 : N. Bergeal et al**

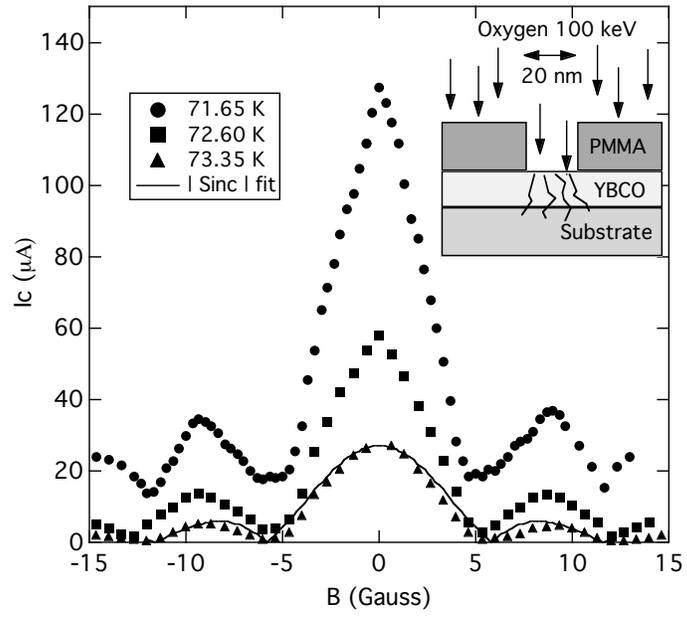

**Figure 2 : N. Bergeal et al**

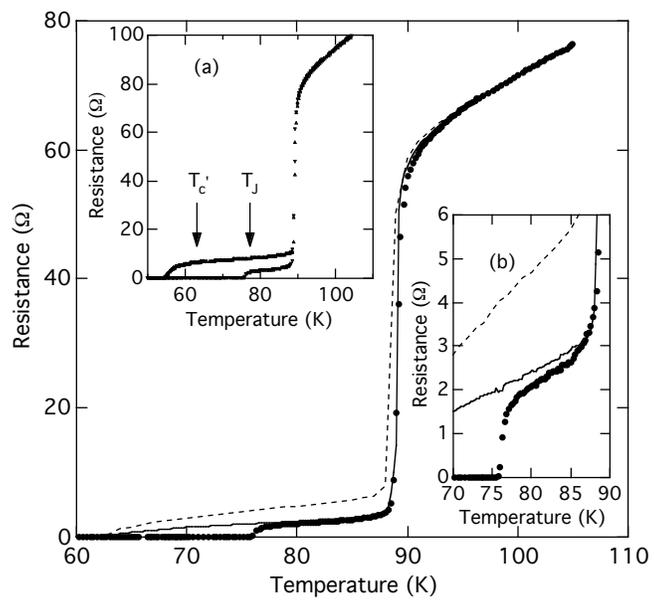

**Figure 3 : N. Bergeal et al**